\documentclass{ws-procs9x6}

\newcommand{\be}{\begin{equation}}
\newcommand{\ee}{\end{equation}}
\newcommand{\bea}{\begin{eqnarray}}
\newcommand{\eea}{\end{eqnarray}}

\newcommand{\nn}{\nonumber}
\newcommand{\ra}{\rightarrow}

\newcommand{\cO}{{\mathcal O}}

\newcommand{\muhad}{\mu_{\mbox{\rm {\scriptsize had.}}}}

\newcommand{\gL}{\frac{1-\gamma_{5}}{2}}
\newcommand{\gR}{\frac{1+\gamma_{5}}{2}}

\begin{document}

\title{Kaon weak interactions in powers of $1/N_c$}

\author{S. Peris}

\address{Grup de Fisica Teorica and IFAE, \\
Universitat Autonoma de Barcelona, \\
08193 Bellaterra (Barcelona) Spain\\}

\maketitle

\abstracts{I review a recent analytic method  for computing matrix
elements of electroweak operators based on the large-$N_c$
expansion. In particular, I give a rather detailed description of
the matching of the quark Lagrangian onto the meson chiral
Lagrangian governing $K^0-\overline{K}^0$ mixing. In the $\Delta
S=1$ sector, I explain how to obtain an estimate for both
$\varepsilon'/\varepsilon$ and the $\Delta I=1/2$ rule. Finally, I
give an example of how the method can also be useful for lattice
calculations.}

\section{Introduction}

The study of kaon weak interactions is difficult due to the fact
that the strong interactions become nonperturbative at a scale
$\Lambda_{QCD}\sim 1$ GeV which is higher than the kaon mass
$M_K$. However, because there is the large hierarchy
$\Lambda_{QCD}/M_W\ll 1$ with respect to the scale of weak
interactions, $M_W$, the problem can be simplified with the use of
Effective Field Theory techniques. In the Effective Field Theory
of the Standard Model valid at the Kaon scale, one deletes from
the Lagrangian any field whose associated particle mass is higher
than $\Lambda_{QCD}$. At the same time one introduces higher
dimension operators with the $u,d,s$ quarks as degrees of freedom
in order to reproduce the processes that in the Standard Model
were taking place due to virtual exchange of the $W$ and other
heavy fields. The introduction of these higher dimension operators
modifies drastically the ultraviolet properties of the Lagrangian
making it look ``nonrenormalizable''. However, if the basis of
higher dimension operators is complete, it is possible to absorb
all the divergences with the counterterms supplied by these
operators. Of course, this takes care of the divergences, but it
leaves a finite piece behind. Furthermore, it is a finite piece
which is dependent on the conventions chosen to do the calculation
(such as, e.g., the regularization scheme chosen --MS or
$\overline{\mathrm{MS}}$--, the precise definition of $\gamma_5$
employed --NDR or HV--, evanescent operators, etc...). Obviously
physical results cannot be scheme/convention dependent and the
ambiguity is resolved in the so-called matching conditions. These
conditions equate a Green's function computed in the Effective
Field Theory and the full Standard Model, imposing that the
physics is the same even though the heavy particles are missing in
the former.

 The advantage of the Effective Field Theory technique
is that it is simpler. This simplicity allows, e.g., the
systematic resummation of all powers of the strong coupling
$\alpha_{s}$ accompanied by large logarithms in combinations such
as $\alpha_{s}^n \log^n M_W/\Lambda_{QCD}$ using renormalization
group techniques\cite{Buras}. However, below ${\Lambda_{QCD}}$
there is no point in resumming powers of $\alpha_{s}$. Fully
nonperturbative effects take place and, in fact, kaon transitions
are described in terms of a Chiral Lagrangian with the kaon field
as an explicit degree of freedom, and no longer in terms of fields
for the $u,d,s$ quarks. The matching condition between the quark
Lagrangian and the Chiral Lagrangian requires a nonperturbative
treatment. This is where the large-$N_c$ expansion comes in. This
expansion is very well suited for this because it can be
implemented both at the level of quarks and at the level of
mesons\cite{largeN}. However, even at the leading order in the
large-$N_c$ expansion, Green's functions in QCD receive a
contribution from an infinity of resonances whose masses and
couplings are unknown. This is why one actually has to resort to
an approximation to large-$N_c$ QCD. This approximation, which has
been termed the ``Hadronic Approximation'' (HA)\cite{MHA},
consists of the ratio of two polynomials whose coefficients are
fixed by matching onto the first few terms in the chiral and
operator product expansions of the Green's function one is
interested in. The rational approximant so constructed constitutes
an interpolator  between the low- and high-momentum regimes, which
one can use to perform the necessary calculations.

\section{$\Delta S=2$ operators and the $K^0-\overline{K}^0$ complex.}
\subsection{Leading contribution: dimension-six quark operators.}\label{subsec:prod}
In the Standard Model, a double exchange of W bosons generates
through the famous box diagram a $\Delta S=2$ transition amplitude
between the $K^0$ and the $\overline{K^0}$. Below the charm mass,
there arises the effective operator
\begin{eqnarray}\label{one}
    &&\mathcal{H}^{S=2}_{eff}\approx\frac{G_F^2}{4\pi^2}
    \ \overline{s}_L\gamma^{\mu}d_L(x)\
    \overline{s}_L\gamma_{\mu}d_L(x)\quad \times \nonumber \\
 && \quad  \ c(\mu)\ \Bigg\{\eta_1 \lambda_c^2 m^2_c+ \eta_2
    \lambda_t^2 \left(m^2_{t}\right)_{eff}+ 2 \eta_3 \lambda_c \lambda_t m^2_c
    \log\frac{m_t^2}{m_c^2}\Bigg\}+ \mathrm{h.c.}\ ,
\end{eqnarray}
where $\psi_{L,R}=\frac{1\mp\gamma_5}{2}\psi,\
\lambda_i=V^{*}_{is} V_{id}$, $m_{t,c}$ are the corresponding
quark masses, $\eta_{1,2,3}$ are some numerical coefficients of
$\mathcal{O}(1)$\cite{Buras2}, and
\begin{equation}\label{three}
c(\mu)= \left(\alpha_s(\mu)\right)^{-\frac{9}{11N_c}}\ \left[1+
\frac{\alpha_s(\mu)}{\pi}\left( \frac{1433}{1936}+ \frac{1}{8}\
\kappa \right)  \right]\
\end{equation}
expresses the running under the renormalization group of the
operator in Eq. (\ref{one}). The parameter $\kappa$ contains the
scheme dependence and equals $0\ (-4)$ in the naive dimensional
regularization (resp. 't Hooft-Veltman) schemes. In the case of
the top one defines the effective mass\cite{Buras2},
\begin{equation}\label{two}
\left(m^2_t\right)_{eff}= 2.39 \ M_W^2 \left[\frac{m_t}{167\
GeV}\right]^{1.52},
\end{equation}
to take into account that the top mass is heavier than the $W$.
Physical observables such as the $K_L-K_S$ mass difference and
$\epsilon_K$ get a contribution from the real and imaginary part,
respectively, of the matrix element \break
$<K^0|\mathcal{H}^{S=2}_{eff}|\overline{K}^0>$.

Below $\Lambda_{QCD}\sim 1$ GeV, it no longer makes any sense to
think of an operator written in terms of quarks and, instead, one
writes a Chiral Lagrangian. The $\Delta S=2$ chiral operator at
order $\mathcal{O}(p^2)$ reads\cite{ed}
\begin{equation}\label{thirteen}
    \mathcal{L}_{eff}^{S=2}= \frac{G_F^2}{16 \pi^2}\ \ F_0^4
    \ \Lambda^2_{S=2} \ \mathrm{Tr}
    \left[\lambda_{32}\left(D^{\mu}U^{\dag}\right)U\lambda_{32}
    \left(D_{\mu}U^{\dag}\right)U\right] + \mathrm{h.c.}\
    ,
\end{equation}
where $[\lambda_{32}]_{ij}=\delta_{i3}\delta_{2j}$ is a (spurion)
matrix in flavor space, $F_0\simeq 0.087$ GeV is the pion decay
constant (in the chiral limit) and $U$ is a $3\times 3$ unitary
matrix collecting the Goldstone boson degrees of freedom, which
transforms as $U\rightarrow RUL^{\dag}$ under a flavor rotation
$(R,L)$ of the group $SU(3)_L\times SU(3)_R$. The scale
$\Lambda^2_{S=2}$ inherits some dependence on short-distance
physics,
\begin{equation}\label{sd}
\Lambda_{S=2}^2|_{\mathrm{dim-6}}=\ \widehat{g}_{S=2}\
\Bigg[\eta_1 \lambda_c^2 m^2_c+ \eta_2
    \lambda_t^2 \left(m^2_{t}\right)_{eff}+ 2 \eta_3 \lambda_c \lambda_t m^2_c
    \log\frac{m_t^2}{m_c^2}\Bigg]\ ,
\end{equation}
but there is also a coupling constant, $ \widehat{g}_{S=2}$, to be
determined via a matching condition.

 What is this matching condition? Since covariant derivatives contain external
fields $l$ and $r$, i.e.
$D^{\mu}U^{\dagger}U=\partial^{\mu}U^{\dagger}U+iU^{\dagger}
r^{\mu}U -il^{\mu}$, one sees that Eq. (\ref{thirteen}) contains a
``mass term'', $r^{\mu}_{\bar d s}r_{\mu}^{\bar d s}$,  for the
external field $r^{\mu}_{\bar d s}$. This $r^{\mu}_{\bar d s}$ is
precisely the external field that couples to the right-handed
current $\bar d_R \gamma^{\mu} s_R$ in the kinetic term for the
quark field in the QCD Lagrangian. Furthermore, a mass term for
the $r^{\mu}_{\bar d s}$ field changes strangeness by two units,
so that it can only come about from the quark Lagrangian because
of the presence of the operator
$(\overline{s_L}\gamma^{\mu}d_L)^2$ in Eq. (\ref{one}).
Consistency demands that the two mass terms be the same. Equating
the term $r^{\mu}_{\bar d s}r_{\mu}^{\bar d s}$ obtained from Eq.
(\ref{thirteen}) to that obtained from Eq. (\ref{one}) (plus gluon
interactions) one obtains the matching condition\cite{BK1}

\begin{eqnarray}\label{GDELTAS2}
\widehat{g}_{S=2}&=& c(\mu)\ g_{S=2}(\mu)\ , \quad \mathrm{where}\nonumber \\
 g_{S=2}(\mu)&=&
 1 - \frac{\mu^2_{had}}{32\pi^2 F_{0}^2}\
\frac{(4\pi\mu^2/\mu^2_{had})^{\epsilon/2}}{\Gamma(2-\epsilon/2)}\!\!
\int_{0}^{\infty}\!\!\!\!\!\!\!dz \ z^{-\epsilon/2} W(z)\,,
\end{eqnarray}
where $\mu^2_{had}$ is an arbitrary scale used to define the
integral in terms of a dimensionless variable $z\equiv
Q^2/\mu^2_{had}$, where $Q$ is a loop momentum. The coupling
$\widehat{g}_{S=2}$ is  renormalization group invariant. The
function $W(z)$ is defined as
\begin{equation}\label{doubleviu}
 W(z)=z\frac{\muhad^2}{F_0^2}
W^{(1)}_{LRLR}(z\muhad^2)\ , \ee where, in turn, $W^{(1)}_{LRLR}$
is defined through an integral over the solid angle of the
momentum $q$, {\be\label{intLRLR} \int d\Omega_{q}\
g_{\mu\nu}W_{LRLR}^{\mu\alpha\nu\beta}(q,l)\vert_{\mbox{\rm
{\scriptsize unfactorized}}}= \left(\frac{l^{\alpha}l^{\beta}}{
l^2}-g^{\alpha\beta}\right)\ W^{(1)}_{LRLR}(Q^2)\, , \ee} and
$Q^2\equiv -q^2$. In Eq. (\ref{intLRLR}) the function
$W_{LRLR}^{\mu\alpha\nu\beta}(q,l)$ stands for { \bea\label{LRLR}
&&W_{LRLR}^{\mu\alpha\nu\beta}(q,l)= \\
&&i^3 \lim_{l\ra 0} \! \int \!\!d^4x d^4y d^4z \ e^{i(qx + ly -
lz)}\langle 0\vert T\{ L_{\bar{s}d}^{\mu}(x)\
R_{\bar{d}s}^{\alpha}(y) \ L_{\bar{s}d}^{\nu}(0)\
R_{\bar{d}s}^{\beta}(z)\}\vert 0\rangle , \nn \eea } with \be
L_{\bar{s}d}^{\mu}(x)=\bar{s}(x)\gamma^{\mu}\gL d(x) \quad , \quad
R_{\bar{d}s}^{\mu}(x)=\bar{d}(x)\gamma^{\mu}\gR s(x)\ . \ee Since
$W(z)\sim \mathcal{O}(N_c^0)$ and $F_0^2\sim \cO(N_c)$ one sees
that Eq. (\ref{GDELTAS2}) yields $g_{S=2}(\mu) = 1-
\cO(N_c^{-1})$, with the unity stemming from the factorized part
of the function $W_{LRLR}^{\mu\alpha\nu\beta}$.

Notice that the function $W_{LRLR}^{\mu\alpha\nu\beta}(q,l)$ is
essentially a 2-point ``left$\times$ left'' Green's function, with
incoming momentum $Q$, with a double insertion of a right-handed
current at zero momentum (i.e. $l\rightarrow 0$). Even though
$W(z)$ is an order parameter of spontaneous chiral symmetry
breaking and, therefore, receives no contribution from
perturbation theory to all orders in $\alpha_s$, the integral in
Eq. (\ref{GDELTAS2}) is divergent (i.e. ill-defined) and must be
regularized. Consistency demands that the regularization used be
the same as in the calculation of the Wilson coefficient
(\ref{three}) and this is why the integral in Eq. (\ref{GDELTAS2})
is done in $\overline{\mathrm{MS}}$ in $4-\epsilon$ dimensions.

 In principle, knowledge of $W(z)$ for
the full range of z is required to calculate the coupling $g_{
S=2}(\mu)$. However, lacking the solution of QCD at large $N_c$,
this information is not known. What is known, nevertheless, is the
low- and high-$z$ expansions of $W(z)$ because they are given by
chiral perturbation theory and the operator product expansion,
respectively. If we build a good interpolator for the region in
between, the answer is ready.

In the large-$N_c$ limit the function $W(z)$ is a meromorphic
function, i.e. it has an infinity of isolated poles on the
negative z axis, but no cut. The Hadronic Approximation (HA) I
shall use is an approximation to this function which consists in
keeping only a finite number of poles, fixing their residues so
that the coefficients of the chiral and operator product
expansions are reproduced. In mathematics, this is called a
rational approximant.

In general the convergence of this type of approximants  is
difficult to establish for an arbitrary function\cite{Bender}.
However, in our case we expect it to work reasonably well. For one
thing we expect $W(z)$ to be a smooth function  since it is a
Green's function in the euclidean. Furthermore, if the operator
product expansion sets in at scales $\gtrsim 1$ GeV and the chiral
expansion sets in at scales $\lesssim M_{\rho}$, then the gap to
interpolate by the approximant is not very large and,
consequently, one may expect the error to be reasonably small. In
particular, we have verified within a model that this
approximation works nicely\cite{phily} and, in fact, the
traditional success of vector meson dominance shows an indication
that things may work similarly for QCD. At any rate, whenever
possible, we shall check on the convergence of the approximation
for the case at hand.

Computing the chiral expansion for $W(z)$ at low $z$ one
finds\cite{chiralbijnens,BK1}
\begin{equation}\label{Wzlowenergy}
W(z)\approx \quad 6 - 24 \ \frac{\mu^2_{had}}{F_0^2}\ \left(2
L_1+5L_2+L_3+L_9\right)\ z\ + \dots \quad  ,
\end{equation}
where the first (resp. second) term comes from the contribution of
$\mathcal{O}(p^2)$ (resp. $\mathcal{O}(p^4)$) in the strong chiral
Lagrangian.

At high $z$ the operator product expansion yields
\begin{equation}\label{Wzhighenergy}
W(z)\approx \frac{24 \pi \alpha_s F_0^2}{\mu^2_{had}}\
\frac{1}{z}\ \left[1 + \frac{\epsilon}{12}\left(5 + \kappa\right)
+  \mathcal{O}(\alpha_s)+ \frac{10}{9}\
\frac{\delta_{K}^2}{\mu_{had}^2}\ \frac{1}{z} \right] + \dots
\quad .
  \end{equation}
The leading term in $1/z$ in this expression stems from the
contribution of dimension-six operators to the operator product
expansion. The $\kappa$ parameter is the same as that in Eq.
(\ref{three}) and contributes a nonvanishing result insofar as the
calculation is done away from 4 dimensions, i.e. $4-d=\epsilon\neq
0$. This is how our matching condition knows about short-distance
renormalization scheme conventions.

\begin{figure}[ht]
\epsfxsize=3cm   
\centerline{{\epsfxsize=1in \epsfbox{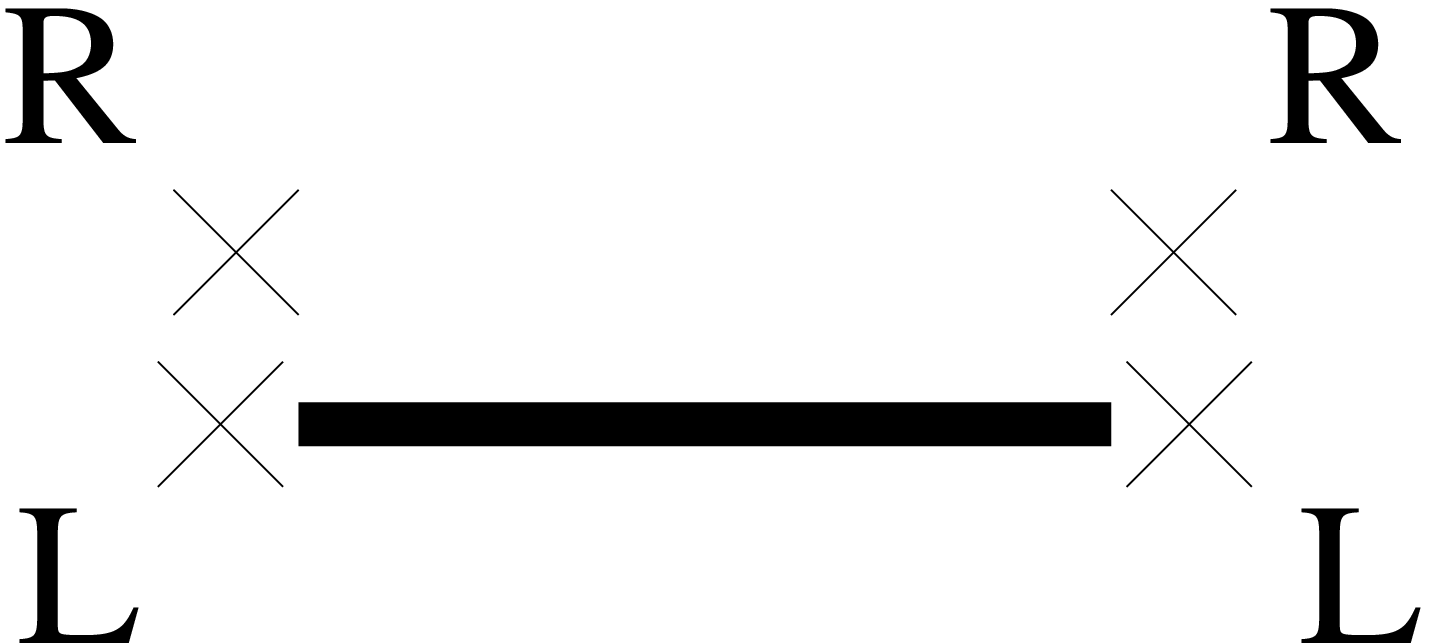}}\hspace {0.3in}
{\epsfxsize=1in \epsfbox{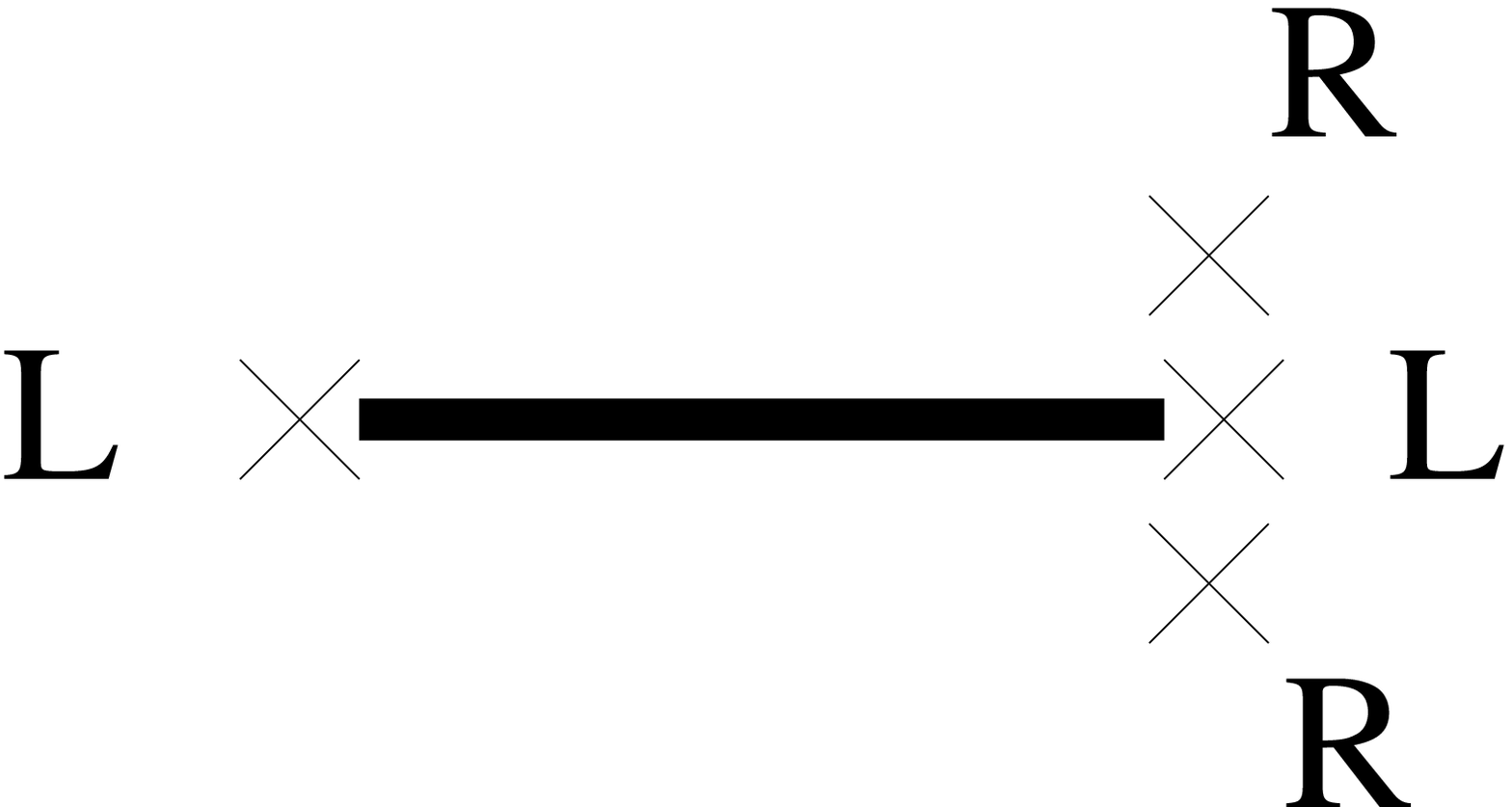}}\hspace{0.3in}
{\epsfxsize=1in \epsfbox{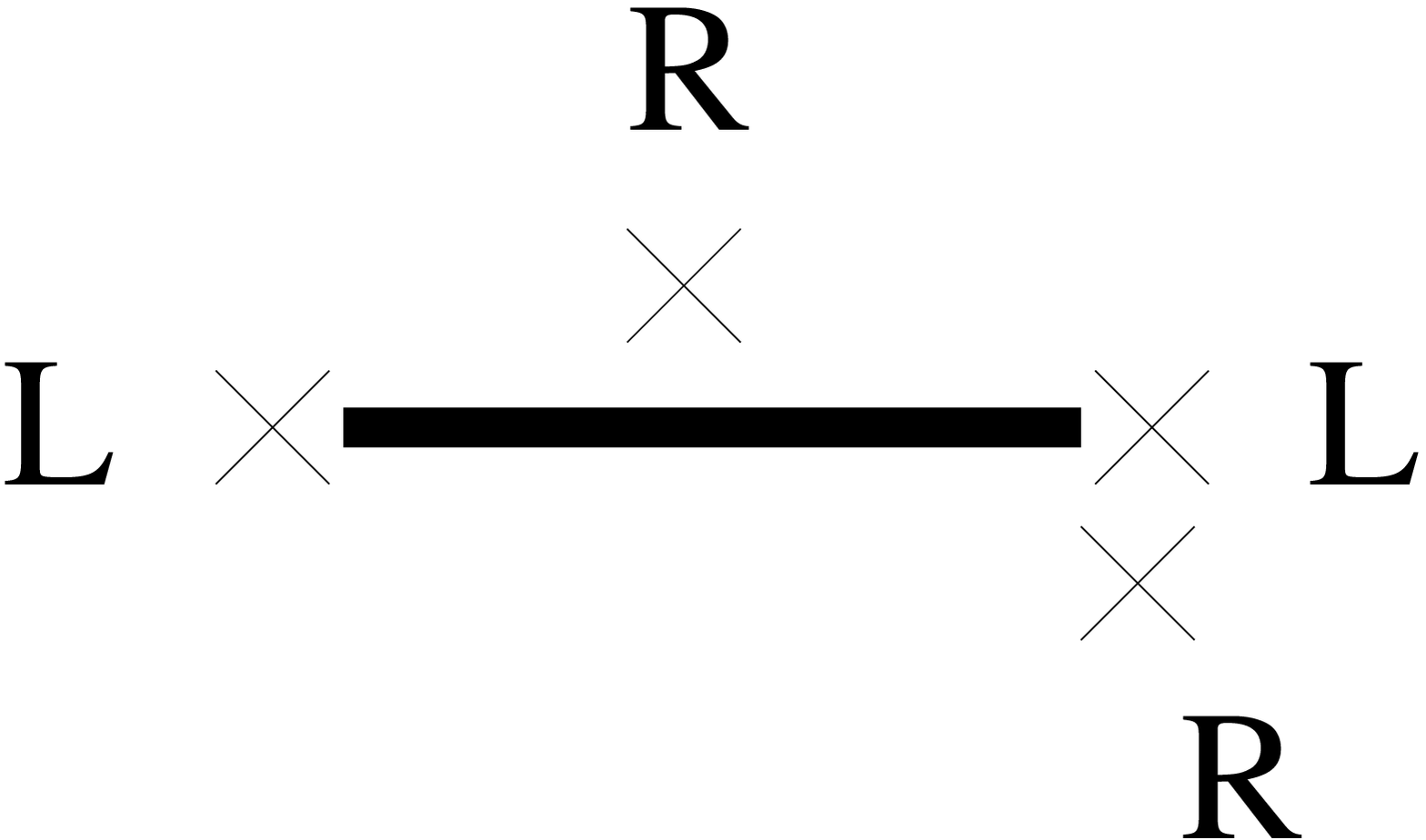}}}
\centerline{\hspace{0.1in} {\epsfxsize=1in
\epsfbox{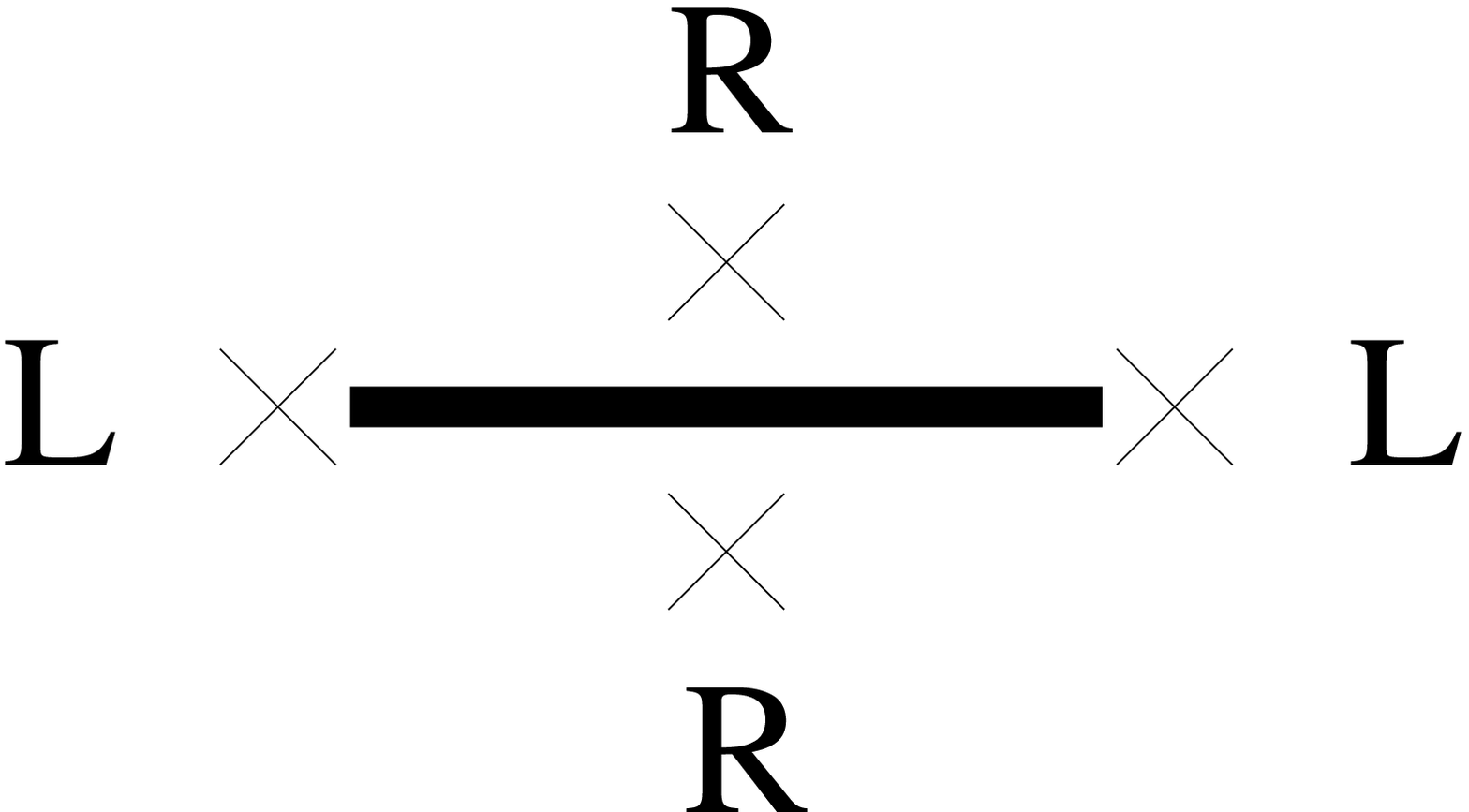}}\hspace{0.3in} {\epsfxsize=1in
\epsfbox{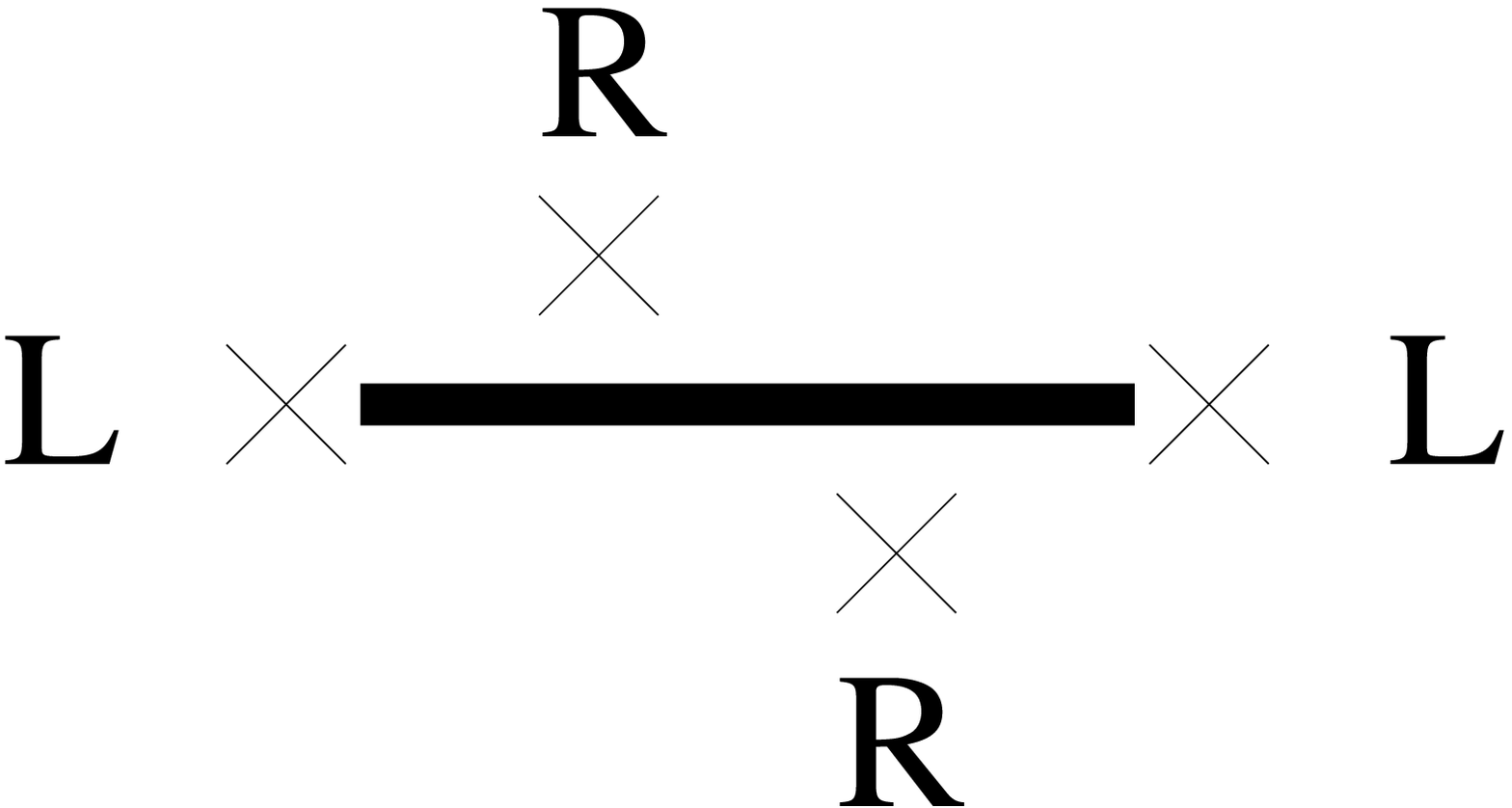}}} \caption{Resonance diagrams contributing
to $W(z)$.} \label{diagrams}
\end{figure}

The $\delta_K$ term in Eq. (\ref{Wzhighenergy}) is the
contribution from dimension-eight operators in the operator
product expansion. We may use this term to monitor the convergence
of our expansion in the following way. First, we may start by
considering the $\delta_{K}$ term in the operator product
expansion as ``subleading'' and neglect its contribution
altogether. After reinstating it later on, and redo the
calculation, we may compare the two results thus obtained.
Clearly, if our approximation is to make sense, both results
should be close to one another.

So let us neglect $\delta_K$ in Eq. (\ref{Wzhighenergy}) for the
time being. Since we wish to calculate the integral in Eq.
(\ref{GDELTAS2}), we need to interpolate between the low- and
high-z regimes in Eqs. (\ref{Wzlowenergy},\ref{Wzhighenergy}). To
this end we observe that the pole structure of the diagrams
depicted in Fig. \ref{diagrams} shows single, double and triple
poles (those which can be obtained by cutting through the solid
line between crosses). In the strict large-$N_c$ limit the
Mittag-Leffler's theorem\cite{Mittag-Leffler} for meromorphic
functions allows us to write $W(z)$ in the form
\begin{equation}\label{Mittag}
    W(z)^{N_c\rightarrow \infty} =\sum_{k}^{\infty}\left\{\frac{a_k}{(z+\rho_k)}+ \frac{b_k}{(z+\rho_k)^2} +
\frac{c_k}{(z+\rho_k)^3}\right\}\quad ,
\end{equation}
where the infinite sum extends over all possible resonances
allowed by quantum numbers. The analytic structure of
(\ref{Mittag}) suggests that, as a first step, we restrict
ourselves to just one resonance in this sum. The HA interpolator
so constructed reads then,
\begin{equation}\label{largeNansatz}
 W(z)=\frac{a_V}{(z+\rho)}+ \frac{b_V}{(z+\rho)^2} +
\frac{c_V}{(z+\rho)^3}\quad ,
\end{equation}
where $\rho= M^2_{\rho}/\mu^2_{had}$, and we determine the
constants $a_V,b_V,c_V$ by imposing that it reproduces the
behavior in Eqs. (\ref{Wzlowenergy}, \ref{Wzhighenergy}). The
result of this interpolation is shown as the dash-dotted curve in
Fig. \ref{plot}. This figure also shows the OPE  and chiral
expansion as dashed curves at large and small values of z,
respectively.

\begin{figure}[ht]
\epsfxsize=3cm   
\centerline{\epsfxsize=3.in \epsfbox{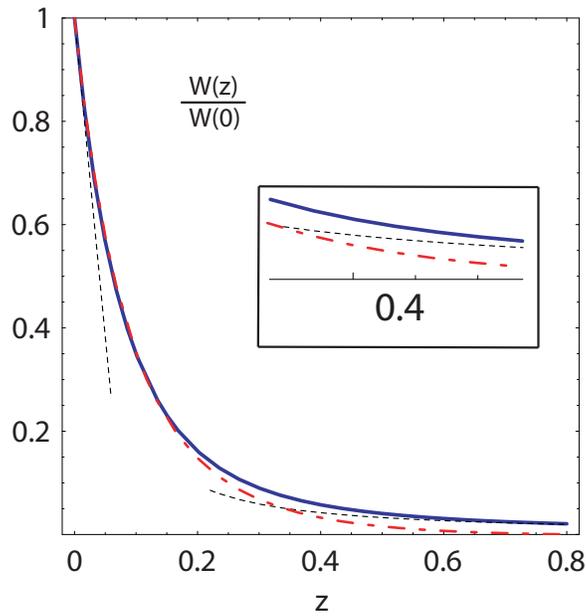}} \caption{The
dash-dotted curve corresponds to the interpolator $W[z]$ in
Eq.~(\ref{largeNansatz}), i.e. without considering the
$\delta_{K}$ contribution, normalized to its value at the origin,
versus $z=Q^2/\mu^2_{had}$ for $\mu_{had}=1.4$ GeV, $F_0=85.3$ MeV
and $M_V= 770$ MeV. The solid curve is the full function $W(z)$ in
Eq. (\ref{VS}) after including the  $\delta_{K}$ contribution in
Eq. (\ref{Wzhighenergy}) (see text below). As one can see both
curves are quite similar (the insert is a blow-up of the region
$z\sim 0.4$). The two dashed curves correspond to the chiral and
operator product expansions, for small and large values of z,
respectively.\label{plot}}
\end{figure}

Notice that the shape of the interpolating function $W(z)$ shown
in Fig. \ref{plot} is very smooth. This kind of shape for the
Green's function is very representative and we have found similar
shapes in all the different cases we have studied.

One can now use the example of $W(z)$ to develop some intuition of
the role played by the different energy scales in the problem.
Although in a one-scale theory like QCD all scales are ultimately
related to each other, it is a fact of life that there is a clear
numerical hierarchy. First, there are chiral parameters such as
$F_{\pi}$ or $\langle\overline{\psi}\psi\rangle^{1/3}$ whose scale
is $\sim 100-200$ MeV. Second, there is a typical resonance mass,
or mass gap, whose scale is much larger, i.e. $\Lambda_{QCD}\sim
1$ GeV. Notice that, since $F_{\pi}\sim N_c^{1/2}$ but
$\Lambda_{QCD}\sim N_c^0$, a naive use of the strict large-$N_c$
limit would lead one to the erroneous conclusion that
$\Lambda_{QCD}/F_{\pi}$ is negligible.

In matching conditions like (\ref{GDELTAS2}) this observation is
crucial. The reason is that the factorized contribution is
governed by $F_{\pi}$ --which in Eq. (\ref{GDELTAS2}) has been
scaled out and this is why the first term is unity-- whereas the
unfactorized contribution is governed by $\Lambda_{QCD}$.

Since $W(z)\sim 1$ in the region $z\lesssim
\Lambda_{QCD}^2/\mu_{had}^2$ --before the OPE sets in-- and
neglecting the logarithmic divergence from the OPE tail, this
means that
\begin{equation}\label{estimate}
    \int_0^{\frac{\Lambda_{QCD}^2}{\mu_{had}^2}} dz\ W(z)
    \sim \frac{\Lambda_{QCD}^2}{\mu_{had}^2}\ ,
\end{equation}
which says that the correction to unity in Eq. (\ref{GDELTAS2})
will be of order $\Lambda_{QCD}^2/(4 \pi F_{\pi})^2$, which is not
a negligible contribution at all. In this case keeping only the
factorized contribution is not safe because the $N_c \rightarrow
\infty $ limit happens to select a scale which is ``abnormally
small'', $F_{\pi}$, as compared to the larger scale
$\Lambda_{QCD}$, which can only show up in the next-to-leading
$1/N_c$ terms. In these cases it is not unnatural to expect large
unfactorized contributions. As a matter of fact, later on we shall
see that the unfactorized contribution is $\sim 50\%$ of the
factorized contribution in the case of $B_K$ but it is several
times larger in the case of the strong penguin contribution to
$\varepsilon'/\varepsilon$.

However, once one has included the unfactorized contribution with
its $\Lambda_{QCD}$ scale, there is no further scale in the game.
Therefore, there is no reason to expect that subleading effects
will still yield larger contributions with the consequent
breakdown of the large-$N_c$ expansion. These subleading effects
are typically either an OZI-violating amplitude or related to
resonance widths, and all evidence so far is compatible with these
two being reasonably small. Notice that something similar to this
happens in the relationship between the $\eta'$ mass and the
topological susceptibility for different number of
colors\cite{Teper}: even though the $\eta'$ mass is not at all
small as compared to $\Lambda_{QCD}$ --whereas it vanishes when
$N_c\rightarrow \infty$-- the $1/N_c$ expansion gives a good
description of the lattice data for different values of $N_c$,
i.e. the expansion does not break down.

Let us come back to our matching condition (\ref{GDELTAS2}).
Having disregarded the $\delta_{K}$ contribution, one sees that
 the interpolator $W(z)$ in Eq.
(\ref{largeNansatz}) crosses the OPE curve at $z\sim 0.4$ (which
is equivalent to $Q\sim 900$ MeV). Merging into the OPE only takes
place at values of $z$ which are larger than those shown in the
plot. Assuming that the OPE is a fair description starting from
$Q\sim 900$ MeV onwards, one can now get an estimate of the
integral in Eq. (\ref{GDELTAS2}) by first integrating with the
interpolator $W(z)$ in the region $0\leq z\lesssim 0.4$ and then
with the OPE curve in the region $0.4\lesssim z < \infty$. The
result is usually presented in the form:
\begin{equation}\label{bk}
    \widehat{B}_K^{\chi}= \frac{3}{4} \ \widehat{g}_{S=2}\ ,
\end{equation}
and one finds $\widehat{B}_K^{\chi}=0.38\pm 0.11$, where the error
is an estimate of higher order $1/N_c$ corrections, $\sim 30 \%$.
This error\cite{BK1} covers all reasonable variations in the input
parameters (e.g. the $\rho$ mass).

 The combination appearing in Eq. ({\ref{bk}) parametrizes
 the $K^0-\overline{K^0}$ matrix element of the
$\Delta S=2$ operator in Eq. (\ref{one}),
\begin{equation}\label{matrixelemnt}
    <\bar{K}^0|\ c(\mu)\ \left(\overline{s}_L\gamma^{\mu}d_L\right)^2(\mu)
    |K^0>= \frac{4}{3}F_{K}^2M_{K}^2\hat B_{K}\quad ,
\end{equation}
\emph{after} including chiral corrections (this is why
$\widehat{B}_K\neq \widehat{B}_K^{\chi}$), and governs quantities
such as $\varepsilon_{K}$ and the $K_L-K_S$ mass difference. The
inclusion of these chiral corrections  is a necessary ingredient
in order to make contact with the physical world since, moreover,
there are indications that they may be sizeable\cite{chiral}. We
plan to be able to report on this in the near future but, for now,
the result after Eq. (\ref{bk}) is still in the chiral limit.

One may now include the contribution from the $\delta_{K}$ term in
the operator product expansion of Eq. (\ref{Wzhighenergy}). This
coefficient $\delta_{K}$ parameterizes the following matrix
element
\begin{equation}\label{delta}
    <0| g_s \bar{s}_L\tilde{G}_{\mu\nu}^a \lambda_a \gamma^{\mu}d_L |K(p)
>=  - i \sqrt{2}F_0\ \delta_{K}^2\ p_{\mu}\ ,
\end{equation}
and a  sum rule analysis\cite{Oscar1} gives  $\delta_{K}^2=0.12
\pm 0.07$ GeV$^2$. Since we now have one more condition at large
$z$, we have to include one more resonance in the interpolator in
order to achieve matching. First, we observe that the low-$z$
behavior of $W(z)$ depends on $L_3$ and this low-energy constant,
unlike the other $L_i$'s appearing in (\ref{Wzlowenergy}),
receives contribution from scalars resonances\cite{swiss}. Second,
a simple pole contribution from a scalar resonance is compatible
with the analytic structure and quantum numbers shown in Fig.
\ref{diagrams}. Therefore, we enlarge the interpolator in Eq.
(\ref{largeNansatz}) with a scalar pole to read
\begin{equation}\label{VS}
     W(z)_{V,S}=\frac{a_V}{(z+\rho_V)}+ \frac{b_V}{(z+\rho_V)^2} +
\frac{c_V}{(z+\rho_V)^3}+ \frac{a_S}{(z+\rho_S)}\quad ,
\end{equation}
and allow for a generous variation of the scalar resonance mass
$M_S=900\pm 400$ MeV. Imposing that $W(z)_{V,S}$ matches both
expansions in Eqs. (\ref{Wzlowenergy}, \ref{Wzhighenergy}) one can
obtain the unknown residues and determine $W(z)$, Eq. (\ref{VS}).
This function is the solid line plotted in Fig. \ref{plot}. One
sees that it matches the OPE in a smoother way than the
interpolator obtained earlier (dash-dotted line in Fig.
\ref{plot}), but there is no dramatic difference in the area
underneath. In summary, the neglect of higher order terms in the
expansions (\ref{Wzlowenergy}, \ref{Wzhighenergy}), as a first
approximation, is a self-consistent procedure\footnote{We have
studied this issue in much more detail in a simple model, with the
same conclusion.\cite{phily}}.

Using now this improved function $W(z)$ and Eq. (\ref{GDELTAS2}),
we can calculate $\widehat{B}_K^{\chi}$ once again and obtain
\begin{equation}\label{grandBK}
    \widehat{B}_K^{\chi}=0.36\pm 0.15\ ,
\end{equation}
as our final value. We emphasize that, because the coupling
$\widehat{g}_{S=2}$ in Eq. (\ref{sd}) is of $\mathcal{O}(p^2)$ in
the weak chiral Lagrangian (\ref{thirteen}), our value is in the
chiral limit. This result is in nice agreement with recent
determinations done on the lattice, where it now begins to be
possible to consider dynamical fermions\cite{lattice}.

\subsection{Corrections from dimension-eight quark operators.}

 The passage from the Lagrangian  in terms of quarks and gluons (\ref{one})
 to the Lagrangian (\ref{thirteen}) in terms of Goldstone mesons requires
 the integration of all the resonances. But in this integration
 the mass scale involved is the mass gap, i.e. $\Lambda_{QCD}\sim 1$ GeV, which is not
 negligible compared to the charm mass. Therefore, in
 the matching condition for $\Lambda_{S=2}^2$
 in Eq. (\ref{sd}) there may be extra contributions of $\mathcal{O}(\Lambda_{QCD}^2)$.

 To analyze the presence of these extra contributions one
 obviously must consider effects
 which are of $\mathcal{O}(\Lambda_{QCD}^2/m_c^2)$, relative to
 the contributions in Eq. (\ref{sd}). Therefore, after integrating out
 the charm quark,
 one must go to a Lagrangian of quark operators
 which are dimension eight rather than dimension six as in Eq.
 (\ref{one}). Since in this case there is not the competition between the scales
 $F_{\pi}$ and $\Lambda_{QCD}$ we discussed earlier, to simplify
 matters, we shall take the large-$N_c$ limit.

Dimensional analysis shows that $\Delta S=2$ dimension-eight
operators, unlike the dimension-six operator in
 Eq. (\ref{one}), cannot come with a quark mass out front. It is
 not surprising, then, that the GIM mechanism  becomes fully
 operational for scales $m_c < \mu < m_t, M_W$ while the charm
 quark is active, arranging combinations like
 $(\lambda_c+\lambda_u)^2=\lambda_t^2$ which is numerically
 negligible since, unlike in the dimension-six case of Eq. (\ref{one}),
 it cannot be compensated by a large $m_t$ factor.

 At $\mu=m_c$, however, the charm quark is integrated out and
 the above mechanism no longer applies. At this scale, therefore,
  $\Delta S=2$ dimension-eight operators do get
 generated. Another simplification occurs, however. Notice that we
 eventually want to match onto the $\mathcal{O}(p^2)$ chiral
 Lagrangian in Eq. (\ref{thirteen}). Therefore any $\Delta S=2$
 operator whose matrix element between a $K^0$ and a
 $\overline{K}^0$ is of higher chiral order is of no interest.
 Furthermore, in  the large-$N_c$ limit four quark operators
 factorize, i.e. they become a product of two color singlets.
 A dimension-eight
operator can be arranged in only two ways: either as a product of
two dimension-four operators, or as a product of a dimension-3
operator times a dimension-5 one. However, the first case yields
contributions of higher chiral order. This is due to the fact that
\begin{equation}\label{current}
    \langle \overline{K}^0(p)
    \vert {\bar{s}}_L\gamma_{\nu}{\mathcal{D}}_{\mu}d_L\vert 0\rangle
     \sim \left( p_{\mu}p_{\nu}- \frac{p^2}{4} g_{\mu\nu}\right)\ ,
\end{equation}
which can be seen by contracting with $g^{\mu\nu}$ and using  the
equations of motion --in the chiral limit--. So, only the
combination $\mathcal{O}(\mathrm{dim-3}) \times
\mathcal{O}(\mathrm{dim-5})$ is possible. However, $\overline{s}
\widetilde{G}^{\mu\nu}\gamma_{\nu}d(x)$ is the \emph{only}
dimension-five current connecting a kaon to the
vacuum\cite{Novikov}. Consequently, this means that there is only
one $\Delta S=2$ dimension-eight operator to consider. This
operator is given by
\begin{equation}\label{eight}
    \mathcal{L}_{eff}= - \frac{G_F^2}{3\pi^2}\ c_8(\mu)\ g_s \overline{s}_L \widetilde{G}^{\mu\nu}\gamma_{\nu}d_L(x)\
    \overline{s}_L\gamma_{\mu}d_L(x)\ + \mathrm{h.c.}\ ,
\end{equation}
where $g_s$ is the strong coupling constant.

Integrating out the charm quark yields for the coefficient $c_8$
the matching condition\cite{oscar2}
\begin{equation}\label{matchcond}
    c_8(m_c) = - \ \frac{7}{6}\ \lambda_u^2 - \frac{13}{6}\ \lambda_c \lambda_u .
\end{equation}
 Below the charm mass the Effective Theory is given simply by
 \begin{equation}\label{twelve-p}
    \mathcal{L}_{eff}=  - \frac{G_F^2}{3 \pi^2}\ c_8(\mu)\
    g_s\overline{s}_L \widetilde{G}^{\mu\nu}\gamma_{\nu}d_L\
    \overline{s}_L\gamma_{\mu}d_L
     - \frac{G_F}{\sqrt{2}} \lambda_u\  \overline{s}_L \gamma^{\mu}u_L\
    \overline{u}_L\gamma_{\mu}d_L + h.c.\ .
\end{equation}
Notice that the second operator is the only one producing $\Delta
S=1$ transitions in the large-$N_c$ limit. Away from this limit
its Wilson coefficient is not unity as in Eq. (\ref{twelve-p}) but
gets corrected by $30\%-40\%$, as a naive estimate of typical
$1/N_c$ corrections would say.

At scales $\mu \lesssim m_c$, while one can still consider QCD in
the perturbative regime, the Wilson coefficient $c_8(\mu)$ runs
due to the fact that the square of the $\Delta S=1$ operator in
Eq. (\ref{twelve-p}), with the two up-quarks propagating in a
loop, mixes into the direct $\Delta S=2$ operator. At lowest order
in the strong coupling constant, which is enough for our purposes,
one finds\cite{oscar2} for scales $\mu \sim \Lambda_{QCD}\sim 1$
GeV,
\begin{equation}\label{twelve}
    c_8(\mu)= c_8(m_c) +
    \lambda_u^2 \log\frac{\mu^2}{m_c^2}\ ,
\end{equation}
where $c_8(m_c)$ is given by Eq. (\ref{matchcond}).

 At scales $\mu \sim M_K\ll \Lambda_{QCD}$
perturbation theory is no longer a valid approximation and the
matching of the Lagrangian (\ref{twelve-p}) to the chiral
Lagrangian (\ref{thirteen}) requires again the machinery of the
Hadronic Approximation to large-$N_c$ we used in the previous
section. Skipping the details of this calculation\cite{oscar2}, we
find that the coupling constant $\Lambda^2_{S=2}$ gets the
following contribution,
\begin{equation}\label{result}
\Lambda^2_{S=2}|_{\mathrm{dim-8}}= \left\{\frac{26}{9}\ \lambda_c
\lambda_u\  \delta_K^2 + \lambda_u^2\ (0.95\
\mathrm{GeV})^2\right\}\left( 1 \pm 0.3\right)\ ,
\end{equation}
where $\delta_K^2$ is the parameter defined in Eq. (\ref{delta}),
and the error is an estimate of the size of a typical $1/N_c$
correction. Notice that all renormalization scale and scheme
dependence has canceled out, as it should. This is one of the
advantages of using the HA framework.

The first contribution in (\ref{result}) bears the mark of charm
in $\lambda_c$ and stems from the matching condition for the
Wilson coefficient $c_8(m_c)$ when charm gets integrated out.
Although it has an imaginary part that contributes to
$\varepsilon_{K}$, its size is governed by the dynamical scale
$\delta_K \sim 350$ MeV, which is small enough relative to $m_c$
to yield a very small correction to $\varepsilon_{K}$.

The second contribution is the result of the running below the
charm mass and subsequent matching onto the chiral Lagrangian
(\ref{thirteen}); this is where resonances get integrated out.
This is why only $\lambda_u$ appears and, also, why the energy
scale, which is a combination of resonance masses and couplings,
is essentially given by the mass gap $\Lambda_{QCD}\sim 1$ GeV.
 Since $\lambda_u$ is purely real it cannot contribute to
$\varepsilon_K$.  The energy scale in (\ref{result}) \emph{is}
comparable to $m_c$, which explains why the $K_L-K_S$ mass
difference gets a sizeable $\sim 10-20 \%$
correction\cite{oscar2}.

\section{$\varepsilon'/\varepsilon$ and the $\Delta I=1/2$ rule: ``penguinology''.}

At scales $\mu \lesssim m_c$ the Standard Model gives rise to 10
four-quark operators capable of producing $\Delta S=1$
transitions\cite{Buras}. All these operators mix as the
renormalization scale evolves.

On the other hand, at the scale of the Kaon, the Effective Field
Theory for $\Delta S=1$ transitions is described by the chiral
Lagrangian\cite{ed}
\begin{equation}\label{ewcl}
\!\mathcal{L}_{eff}^{\Delta S=1}=-\frac{G_F}{\sqrt{2}}\
V_{ud}V_{us}^*  \left\{g_{8}\mathcal{L}_{8}+
g_{27}\mathcal{L}_{27}+e^2 g_{ew}\mathrm{Tr}\left(U\lambda_
{L}^{(32)}U^{\dag}Q_R \right) \right\} \,,
\end{equation}
where
\begin{equation}
\mathcal{L}_8=\sum_{i=1,2,3}(\mathcal{L}_{\mu})_{2i}\
(\mathcal{L}^{\mu})_{i3}\quad ,\quad
\mathcal{L}_{27}=\frac{2}{3}(\mathcal{L}_{\mu})_{21}\
(\mathcal{L}^{\mu})_{13}+ (\mathcal{L}_{\mu})_{23}\
(\mathcal{L}^{\mu})_{11}\, ,\nonumber
\end{equation}
 with $\mathcal{L}_{\mu}=-iF_{0}^2\ U(x)^{\dagger}D_{\mu} U(x)$
 , $\lambda_{L}^{(32)}=\delta_{i3}\delta_{j2}$ and
$Q_{L,R}=\mathrm{diag}(2/3,-1/3,-1/3)$.

In this section we shall concentrate on the contribution to
$\varepsilon'/\varepsilon$ from $Q_6$  which, together with the
one from $Q_8$\cite{Knecht,ewothers}, is the dominant
one\cite{Jamin}. As we shall see, the analysis will tell us
interesting things about the $\Delta I=1/2$ rule as well.

We may now run the Effective Field Theory from $M_W$ down to the
charm mass as usual but, at this point, make some simplifications
in the analysis not to have to deal with the full $10\times 10$
operator mixing matrix. Our first simplification will be to stay
within the leading-log approximation.  Furthermore we shall go to
first subleading order in the $1/N_c$ expansion but keeping only
those terms which are enhanced by an extra factor of $n_F$, the
number of flavors. In this case, the operator $Q_6$ only mixes
with $Q_4$, where
\begin{equation}\label{penguins}
  Q_6 = -8\sum_{q=u,d,s}
  \left(\overline{s}_Lq_R\right)\left(\overline{q}_Rd_L\right)\quad , \quad
  Q_4 = 4 \sum_{q=u,d,s}
  \left(\overline{s}_L\gamma^{\mu}q_L\right)\left(\overline{q}_L\gamma_{\mu}d_L\right)\ ,\
\end{equation}
and the sum over color indices within brackets is understood.

The matching condition then reads\cite{HPdeR}
\begin{eqnarray}\label{matching2}
g_{8}(Q_4,Q_6)\!\! &=&\!c_{6}(\mu)
\Bigg\{\frac{-16L_{5}(\mu)\langle
\overline{\psi}\psi\rangle^2(\mu)}{F_{0}^6} + \frac{8n_f}{16\pi^2
F_{0}^4}\!\!\int_{0}^{\infty}\!\!\!\!\!dQ^2 Q^{2}
\mathcal{W}_{DGRR}(Q^2)
\Bigg\}_{\!\!\!\overline{\mathrm{MS}}} \nonumber \\
  &+& c_{4}(\mu)\ \Bigg\{1 - \frac{4n_f}{16\pi^2 F_{0}^4}
 \int_{0}^{\infty}\!\!\!\!\! dQ^2 Q^{2}\ \mathcal{W}_{LLRR}(Q^2)
\Bigg\}_{\overline{\mathrm{MS}}}\ .
\end{eqnarray}
In the previous expression the subscript $\overline{\mathrm{MS}}$
is a reminder that these integrals are UV divergent and have to be
regularized and renormalized using the same scheme as for the
Wilson coefficients $c_{4,6}$. This is also true for $\langle
\overline{\psi}\psi\rangle$ and $L_5$. As far as the $N_c$
counting goes, all the unfactorized contributions, which are of
$\mathcal{O}(n_F/N_c)$, are contained in the terms proportional to
the functions $\mathcal{W}_{DGRR}$ and $\mathcal{W}_{LLRR}$. The
terms proportional to $L_5$ and unity correspond to the factorized
contribution from $Q_6$ and $Q_4$ --respectively-- and, formally,
are of $\mathcal{O}(N_c^0)$.

The functions $\mathcal{W}_{DGRR}$ and $\mathcal{W}_{LLRR}$ are
defined through the connected four-point Green's functions
\begin{eqnarray}\label{green}
  && \!\!\!\!\! \!\!\!\!\!\!\!\!\!\!\! \mathcal{W}_{DGRR}^{~~~~\alpha\beta}(q)=
  i^3 \!\!\! \int \!\! d^4x\ d^4y\ d^4z\
e^{i q.x} \langle T\{ D_{\bar{s}q}(x)G_{\bar{q}d}(0)
R_{\bar{d}u}^{\alpha}(y)R_{\bar{u}s}^{\beta}(z)\}\rangle\ ,\nonumber \\
&& \!\!\!\!\!\!\!\!\!\!\!\!\!
\!\!\mathcal{W}_{LLRR}^{\,\mu\,\mu\,\alpha\,\beta}(q) = i^3\!\!\!
\int \!\! d^4x\ d^4y\ d^4z\ e^{iq\cdot x} \langle T\{
L_{\bar{s}q}^{\mu}(x)L^{\bar{q}d}_{\mu}(0)
R_{\bar{d}u}^{\alpha}(y)R_{\bar{u}s}^{\beta}(z)\}\rangle\ ,
\end{eqnarray}
after integration over the solid angle in $q$-momentum space, as
in Eq. (\ref{intLRLR}). In these expressions
$D_{\bar{s}q}=\overline{s}_Lq_R,\ G_{\bar{q}d}=\overline{q}_R
d_L,\ L_{\bar{s}q}^{\mu}=\overline{s}_L\gamma^{\mu}q_L, \
R_{\bar{d}u}^{\alpha}=\overline{d}_R\gamma^{\alpha}u_R$, etc. It
is not a coincidence that the pair of fermion bilinears which make
up the operators $Q_{6,4}$ in Eq. (\ref{penguins}) also appear in
these Green's functions, although they are located at different
space-time points. It is the integral over $Q$ in Eq.
(\ref{matching2}) which puts these two points back on top of each
other. The matching condition (\ref{matching2}) imposes that the
same ``mass term'' $r_{\overline{d}u}^{\alpha}
r^{\overline{u}s}_{\alpha}$ is obtained when computed from the
covariant derivatives in the Chiral Lagrangian (\ref{ewcl})
--which yields directly $g_8$-- as when computed from the
four-quark Effective Lagrangian, which requires the insertion of
the combination  $c_6 Q_6 + c_4 Q_4$ in the form shown in Eq.
(\ref{matching2}).

In order to calculate the unfactorized contribution one should now
construct the Hadronic Approximation. As in the previous case of
$B_K$, an explicit cancelation of the renormalization scale
dependence is achieved. This is good news since the scale
dependence of the factorized contribution is very large: it
changes by a factor $\sim 2$ if the renormalization scale is
varied in the range $M_{\rho}\leq \mu \leq 1$ GeV. Since this
dependence has to be canceled by the unfactorized contributions,
it is not unthinkable that these contributions be large. As
discussed in the previous section on more general grounds, this
could even be expected.

And, indeed, it was found\cite{HPdeR} that the unfactorized
contribution from the operator $Q_6$ to the coupling $g_8$ in the
matching condition (\ref{matching2}) was a factor $\sim 3$ larger
than the factorized contribution at a scale $\mu =0.8$ GeV.
Although the effect was somewhat smaller for $Q_4$, an enhancement
was found there as well. Both large contributions come from what
in the jargon is called an ``eye'' diagram (i.e. the contraction
of the ``dummy'' quark $q$ in the sums in Eq. (\ref{penguins})).
On the basis of a Nambu-Jona-Lasinio model, this enhancement has
also been found by Bijnens and Prades\cite{Bijnens}.

Decomposing the Wilson coefficients as\cite{Buras}
\begin{equation}\label{wilson}
    c_i(\mu)\!=z_i(\mu)+\tau y_i(\mu)\quad ,\quad
    \tau \!=-\frac{V^*_{ts}V_{td}}{V^*_{us}V_{ud}}\ ,
\end{equation}
the imaginary part of the coupling constant $g_{8}$ in Eq.
(\ref{matching2}) becomes
\begin{equation}\label{img8}
    \mathrm{Im}g_{\underline{8}}\simeq (3\pm 1)\times \mathrm{Im}\tau
     \ ,
\end{equation}
where the error is the result of varying the quark condensate
(which is the source of the biggest uncertainty), in the range
$\langle\overline{\psi}\psi\rangle^{1/3}(2\
\mathrm{GeV})=(0.240-0.260)$ GeV. As to $\mathrm{Im}\tau$, its
current value\cite{Buras2} is $-6.0(5)\times 10^{-4}$.

We can also estimate $\mathrm{Re}g_8$ and $g_{27} $. This is a
very tough test for any calculational framework, as it is in these
parameters that the $\Delta I=1/2$ rule is rooted; a rule which
still to this date defies detailed understanding. However, at the
scale $\mu =m_c$, the situation is particularly simple since only
the operators
\begin{equation}\label{q12}
Q_2 =4(\bar{s}_{L}\gamma^{\mu}u_{L})
(\bar{u}_{L}\gamma_{\mu}d_{L})\quad , \quad
Q_1=4(\bar{s}_{L}\gamma^{\mu}d_{L})(\bar{u}_{L}\gamma_{\mu}u_{L})\
\end{equation}
have nonvanishing real Wilson coefficients. Although at scales
$\mu \ll m_c$ penguin operators will again come into play, it is
not crazy to neglect this effect and stay ``as if $\mu=m_c$'' in a
first approximation. In this case the contribution to
$\mathrm{Re}g_8$ from $Q_{1,2}$ can be estimated because the
non-eye diagrams are related to those appearing in the matching
condition for $\widehat{g}_{S=2}$ in the case of
$B_K$,\cite{pichdeR}  while the eye diagram of $Q_2$ can be
estimated from the eye diagram of $Q_4$ in the matching condition
(\ref{matching2}), just setting $n_F=1$. In numbers this leads
to\cite{HPdeR} $\mathrm{Re}g_{8}=2.1\pm 0.8$, to be compared to
the experimental result\cite{Ecker,Hans} $\mathrm{Re}g_{8}=3.6 \pm
0.1$, after subtraction of chiral corrections. In spite of the
large errors involved I find this result quite encouraging, mainly
because the strict large-$N_c$ result (factorization) would lead
to $\mathrm{Re}g_{8}=0.6$ which is way too small. We begin to see
the large unfactorized contributions which are indispensable for
understanding the $\Delta I=1/2$ rule although, clearly, a more
detailed analysis is needed before victory can be claimed.

$SU(3)$ allows to relate $\widehat{g}_{S=2}$ in $B_K$ to $g_{27}$
because their corresponding matching conditions to the four-quark
Lagrangian are $SU(3)$ rotations of each other\cite{Donoghue}.
This leads to $g_{27} = \widehat{g}_{S=2}= 0.27 \pm 0.12$, to be
compared to  $g_{27} = 0.30 \pm 0.01$ as extracted from
experiment\cite{Ecker,Hans}.

Using all these results, together with\cite{Knecht,ewothers}
\begin{equation}\label{gew}
    \mathrm{Im}\left(e^2g_{ew}\right)=
    (1.6 \pm 0.4)\cdot 10^{-6}\ \mathrm{GeV}^{6} \times \mathrm{Im}\tau\
    ,
\end{equation}
and, including final state interactions\cite{PPS} and isospin
breaking effects\cite{Ecker}, one obtains\cite{HPdeR}
$\varepsilon'/\varepsilon= (5\pm 3) \times 10^{-3}$ to be compared
to the experimental number\cite{exp} $(1.66 \pm 0.16)\times
10^{-3}$. This is not a bad result considering our approximations.

\section{Application to lattice QCD: quenching penguins.}

Current calculations of $\varepsilon'/\varepsilon$ done on the
lattice require getting rid of the fermion determinant in the path
integral in order to be numerically efficient. This is
accomplished by introducing some ghosts quarks which, although
spin 1/2 particles, commute. This lattice technique, known as
``quenching'', has some dramatic consequences. In particular the
flavor symmetry group is changed from the usual $SU(3)_L\times
SU(3)_R$ to a graded $SU(3|3)_L\times SU(3|3)_R$\cite{Bernard}.
Furthermore, and even more importantly, there is no reason for the
quenched theory to have the same weak low-energy constants as the
true theory. The fact is that the current result for
$\varepsilon'/\varepsilon$ on the lattice is 3 times smaller than
the experimental result, and with the opposite sign\cite{RBC}.

A particularly clear example of this is the transformation
undergone by the strong penguin operator $Q_6$ in the quenched
theory\cite{Golterman}. After quenching, the operator $Q_6$ in Eq.
(\ref{penguins}) is no longer a singlet under $SU(3)_R$, or rather
under $SU(3|3)_R$. Instead, it can be decomposed as
\begin{equation}\label{qchq6}
Q_6=\underbrace{2(\overline{s}_L^\alpha\gamma_\mu d_L^\beta)
(\overline{q}_R^\beta\gamma^\mu
q_R^\alpha+\overline{\widetilde{q}}_R^\beta\gamma^\mu
\widetilde{q}_R^\alpha)}_{\frac{1}{2}Q_6^{QS}}+
\underbrace{2(\overline{s}_L^\alpha\gamma_\mu d_L^\beta)
(\overline{q}_R^\beta\gamma^\mu
q_R^\alpha-\overline{\widetilde{q}}_R^\beta\gamma^\mu
\widetilde{q}_R^\alpha)}_{Q_6^{QNS}}
\end{equation}
where the tilde on top refers to the ghost quarks. The flavor
properties are such that $\mathcal{Q}_6^{QS}$ is a singlet under
the quenched flavor group $SU(3|3)_R$ whereas
$\mathcal{Q}_6^{QNS}$ is not. This changes their chiral
representation accordingly\cite{Golterman}:
\begin{eqnarray} \label{bosonization}\mathcal{Q}_6^{QS}&\Rightarrow&
-\ \alpha^{(8,1)}_{q1}\ \mathrm{STr}\left(\Lambda \mathcal{L}_\mu
\mathcal{L}^\mu\right)+\ \alpha^{(8,1)}_{q2}\
\mathrm{STr}\left(2B_0\Lambda(U
M+MU^\dagger)\right)\nonumber \qquad \\
 \mathcal{Q}_6^{QNS}&\Rightarrow&
f^2\alpha^{NS}_q\ \mathrm{STr}\left(\Lambda U\widehat{N}
U^\dagger\right)\ ;\ \widehat{N}=\frac{1}{2}\,{\rm
diag}\left(1,1,1,-1,-1,-1\right)\qquad
\end{eqnarray}with $\overline{q}\Lambda q=\overline{s} d$, $M$ the quark-mass
matrix, and $\mathrm{STr}$ is the so-called supertrace.
$\widehat{N}$ exhibits the non-singlet structure of
$\mathcal{Q}_6^{QNS}$. Comparing Eqs. (\ref{ewcl}) and
(\ref{bosonization}), one sees that the couplings
$\alpha^{(8,1)}_{q1,2}$ have a counterpart in the true theory (the
weak mass term of $\alpha^{(8,1)}_{q2}$ is not written in Eq.
(\ref{ewcl})), but $\alpha^{NS}_q$ is a total quenching artifact.

As we did in previous sections, it is now straightforward to apply
large $N_c$ to determine these coupling constants
$\alpha^{(8,1)}_{q1},\ \alpha^{NS}_q$. One conclusion follows
immediately: because of the presence of the degenerate ghosts
quarks, the sum over flavor in the internal propagator of the eye
diagrams exactly vanishes. But this contribution was actually the
dominant one in the QCD case!. Thus one obtains\cite{GP}:
\begin{itemize}
    \item the contribution from $Q_6$ to
$\alpha^{(8,1)}_{q1}$ is much smaller than its counterpart in the
true theory, $g_8$; and,
    \item  $\frac{\alpha_{q}^{NS}}{\alpha_{q1}^{(8,1)}}= \frac{1}{16 L_5}\simeq 60\ $,
\end{itemize}
where the quenched result\cite{sommer} $L_5\simeq 10^{-3}$ has
been used. A recent lattice analysis\cite{Laiho} has obtained
\begin{equation}\label{laiho}
    \frac{\alpha_{q}^{NS}}{\alpha_{q1}^{(8,1)}}=57(13)\ ,
\end{equation}
in nice agreement with our prediction (not postdiction!) above.

\section{Conclusions.}

Nature has an approximate chiral symmetry because the $u,d,s$
quark masses are small. On a lattice, Nature is approached from
the side of heavy quark masses by using chiral symmetry to guide
the extrapolation. However, chiral symmetry is a property which is
very difficult to achieve on the lattice, taking long hours of
calculations with sophisticated codes and expensive computers and,
regretfully, this becomes an important source of error.

Large-$N_c$ QCD offers the possibility to approach the problem
from the other end.  In the continuum chiral symmetry is much
easier to achieve: it only takes the time needed to write a chiral
Lagrangian. Furthermore, analytic calculations yield an
understanding of the problem which allows building physical
intuition. However, the flip side is that chiral symmetry is only
exact when the quark masses are zero. To get to the real world,
one must work one's way up to realistic quark masses by computing
chiral corrections.

I hope that the work presented here shows, among other things, how
the continuum large-$N_c$ expansion may complement the lattice
approach towards understanding kaon weak interactions.

\section*{Acknowledgments}
I thank N. Scoccola, J. Goity, R. Lebed, A. Pich and C. Schat for
their kind invitation to the ``Workshop on Large-$N_c$ QCD 2004''
where this material was delivered, and for the organization of
this interesting meeting.

I also thank M. Golterman, T. Hambye, M. Knecht, M. Perrottet and
E. de Rafael for a very enjoyable collaboration and countless
interesting discussions, and M. Golterman and E. de Rafael for
their comments on the manuscript.
 This work has
been supported in part by TMR, EC-Contract No. HPRN-CT-2002-00311
(EURIDICE) and by the research projects CICYT-FEDER-FPA2002-00748
and 2001-SGR00188.

%
%
%
%

\end{document}